\documentclass{kapedbk} 

\usepackage[dvips]{graphicx}

\chaptersection 

\upperandlowercase

 \let\footnote\savefootnote

\let\footnotetext\savefootnotetext

\kluwerbib 

\begin{document}

\articletitle{How are AGN Found?}

\rhead{How are AGN Found?}

\author{Richard Mushotzky}

\affil{NASA/Goddard Space Flight Center \\
Code 662, Greenbelt, MD 20771, USA\\}
\email{mushotzky@milkyway.gsfc.nasa.gov}

\begin{abstract}
We discuss the very different methods in each wavelength band for 
selecting and finding Active Galactic Nuclei (AGN). We briefly review 
the history of the different techniques for finding AGN and compare 
and contrast the advantages and difficulties of selection in different 
wavelength bands. We stress the strong selection effects in each 
wavelength band and the difficulty of defining complete samples. Of 
all the techniques presently used, we conclude that selection in the 
hard X-ray band via imaging and spectroscopy is the most complete 
and allows the best estimate of the number and evolution of active 
galaxies. However, all of the techniques have difficulties at low 
luminosities where emission due to stellar processes can have similar 
sizes and luminosities.
\end{abstract}

\section{Introduction}

Looking at the 60 year history of observations of active galaxies, it
is clear that the definition of what they are has strongly influenced
the methods of finding them. From our present perspective, many of the
techniques used over the past 40 years are not truly appropriate and
are more in the line of the famous joke of the drunk looking under
the lamp post for his lost car keys. In this chapter,
I will use the words Active Galactic Nuclei (alias AGN or quasars)
to be the equivalent of radiating supermassive black holes, even 
though this perspective is very recent.

The difficulty in finding AGN is defining what makes the 
observed\footnote{While very frequently the inferred emitted radiation 
is rather different from that intrinsically produced by the region around 
the black hole, the nature of surveys is such that we must rely on the
observed properties of these sources in order to find and identify
them.} radiation different from that due to other processes, in particular
those related to normal stars and stellar evolution 
(e.g., supernovae).\footnote{To date, all the surveys for AGN have relied 
on detection of radiation across the electromagnetic spectrum. Perhaps 
in the distant future we will be able to search for AGN via neutrinos,
gravitational waves, or even very high-energy cosmic rays, but this is
still quite uncertain.} This has often been a process
of exclusion; that is, the emission does not resemble
that from stars or stellar processes. Dust, high-luminosity 
emission from starbursts, and the possible effects of
unusual types of stars complicate the issue. The strong effects of
observing in different spectral ranges also need to be taken into
account; for example, at $R=22$ there are only 100 ``optically-selected'' 
AGN per square degree, but at the equivalent flux level in the $2-10$~keV
X-ray band of $3\times 10^{-15}$~ergs~cm$^{-2}$~s$^{-1}$, there are 
1000~deg$^{-2}$.
Finally, it is clear that the ``non-stellar'' signature has a
wide variety of forms that gives rise to the ``zoo'' of names for 
active galaxies. The spectral energy distributions, optical 
emission-line properties (strengths, widths, and nature), 
line of sight column
densities, time variability characteristics, and bolometric
luminosities of Seyfert~1 galaxies, Seyfert~2 galaxies, 
BL Lacertae objects, LINERs (Low-Ionization Nuclear Emission Regions),
and quasars (to use the names of the largest samples of objects) are 
all rather different (see, e.g., Risaliti \& Elvis, this volume).

It has taken
many years and a large amount of effort to finally come to the
realization that all these classes are manifestations of the same
underlying physical process---emission from near to a supermassive black
hole. However, even today it is not certain if all of these sources
are driven solely by accretion, or whether there is also energy
extraction from the spin of the black hole (see Armitage, this volume). 
It is also not clear if
the energy production is dominated by radiation, relativistic
particle production, or bulk motion of material. It is entirely
possible that there is a simple relation between the ``names'' of the
sources and their physical natures, but at present this seems very
complex and not unique. As opposed to stellar classifications, there is
not a one-to-one relationship between the class of the object and its
physical nature.  However, there are some clear distinctions: for
example, in BL Lacertae objects, the observed radiation is dominated by
emission from relativistic particles in a jet in our line of sight,
and in Seyfert~2 galaxies, the line of sight to the central source 
is blocked by large amounts of dust and gas (see Hewett \& Foltz 1994 
for an earlier review and a detailed discussion of the many systematic
effects in quasar surveys).

Before discussing the field in general, it is important to consider
what a survey really is. As Hewett \& Foltz (1994) point out, there
are three types of surveys: (1) those that find objects, (2) those that
find objects consistently, and (3) those with well-defined selection 
criteria that allow probabilities to be assigned for selection as a 
function of survey parameters. 
Surveys of the first type are the easiest, since the goal is only
to provide sources for study that meet some criteria. Surveys of
the second type are homogenous in their properties, but completeness 
is not important. Many of the issues discussed below are more or less
important depending on which type of survey is being performed. It is
only surveys of the third type that allow comparisons to be made of 
different wavelength regimes and different survey techniques. While 
these surveys are the most scientifically important, they are also
the most difficult to do.

\subsection{A Short History of AGN Search Techniques}

From a historical perspective (e.g., Osterbrock 1991), the
first indications of ``non-stellar'' activity in the nuclei of galaxies
came from the discovery of strong, broad emission lines in NGC1068
(Fath 1913; Slipher 1917) and the discovery of the jet in M87 (Curtis
1917). The spectral features found in NGC1068 are almost never found
in stars or supernovae remnants and thus were an indication of some
new phenomena occurring. It took 50 years before the morphology of
the ``jet'' in M87 was related to non-stellar processes.\footnote{We
now know that jets can also occur in stellar processes (for
example, in Herbig-Haro objects, young neutron stars [e.g., see
the {\em Chandra\/} image of the Vela pulsar], supernovae remnants 
[e.g., see the {\em Chandra\/} image of Cassiopeia A], and the X-ray 
and radio emission from some luminous galactic X-ray sources 
[e.g., Cygnus X-3]). It seems as if the 
jet phenomenon is related to the emission of large amounts of 
energy over a short period of time and into a restricted volume.}
The first ``sample'' of non-stellar activity was that of Seyfert in 1943, 
who found a wide variety of strong ``broad'' lines in the nuclei---but 
not elsewhere---of several otherwise ``normal'' galaxies. It was clear 
from this early paper that there was something quite unusual about 
these sources and that they must be fairly common, but it took almost 
20 more years for significant progress to be made.

This next step occurred with the discovery of extragalactic radio
sources and attempts to find optical counterparts for them. The
discovery of low to moderate redshift ``active galaxies'' as the
optical counterparts to several bright radio sources (Baade \&
Minkowski 1954) showed a new type of ``active galaxy''. It was clear
even in 1958 (Minkowski) that there was tremendous scatter in
the optical properties of the identifications at a fixed radio flux.
Schmidt (1963), Greenstein \& Matthews (1963), and
Schmidt \& Matthews (1964) discovered that the optical counterparts 
of several luminous radio sources were stellar-looking sources at 
large redshifts, and thus were very luminous, compact, extragalactic
sources with non-stellar spectra. They were subsequently named
``quasars'' for quasi-stellar radio sources. The optical properties of
these radio sources were very similar to each other, indicating that
a class of objects had been found. Sandage (1965) realized that there
were sources with the same general optical properties as quasars 
that were not radio sources. These sources had blue colors, meaning a 
large ultraviolet (UV) flux relative to the classical optical band, 
fairly high variability in their continuum intensity, and most had 
strong, broad emission lines over a wide range of ionization.

It was rapidly realized that the nuclei of some ``Seyfert galaxies''
had similar properties to quasars (Woltjer 1959; Burbidge, Burbidge,
\& Sandage 1963), and for the last 30 years, these sources have been
grouped under the name Active Galactic Nuclei or AGN (I think that
the first use of this name in the literature is from Burbidge 1970).
However, even early on, it was clear that not all AGN resembled
quasars. There were Seyfert~2 galaxies, which do not have broad
lines or strong non-stellar continua but do have strong, narrow forbidden
lines that could not be produced by ionization from normal stars. Also,
there were BL Lacertae objects, which usually do not have optical or UV
emission lines but do have a very strong non-thermal continuum and 
a wide variety of optical colors. It was thus clear from the
start that identifying complete samples of AGN would require a 
wide variety of techniques and criteria.

Outside of trying to identify the optical counterparts to the radio
sources (it took over 30 years to completely identify all of the sources
in the 3CR survey, the first large radio survey; Spinrad et al.\ 1985), 
the first systematic search for AGN that I can find in the literature was 
the realization (Arp et al.\ 1968) that the Markarian survey of compact
galaxies with blue stellar colors contained a large number of sources
with the properties of Seyfert galaxies. At almost the same time,
Sargent (1970) used similar critieria for sources from the Zwicky
survey and also found numerous AGN. These two early works, combined
with the radio surveys, set the standard for AGN surveys: using
photographic techniques to find sources with compact, blue nuclei and
following up with optical spectroscopy (see Weedman 1977 for an early
review). However, these searches were completely empirical; i.e.,
they were looking for sources that had the properties of sources
that one already knew were active galaxies.

It is clear that AGN have a very wide range of relative parameters,
from the line strengths and line widths (ranging down to sources
without any emission lines at all; e.g., BL Lacertae objects), to the 
continuum colors, to the amplitudes and timescales of variability. 
Thus, having an inclusive definition is very difficult. It is fair to 
say that most workers have had a difficult time coming up with 
an AGN definition
that is totally complete and not subject to noise and strong selection 
effects. Recent large optical surveys, such as the Two Degree Field (2dF) 
and the Sloan Digital Sky Survey (SDSS), have focused on well-defined 
color or line strength criteria that allow them to be well defined but 
clearly incomplete.

\section{AGN Spectral Energy Distribution and How One Finds AGN}

By definition, emission from a black hole does not resemble that from
an ensemble of normal stars. It has a different broadband spectral
shape (the Spectral Energy Distribution or SED), a very high-luminosity
density, and very different time variability properties. Roughly
speaking, the broadband spectrum of an optically-selected AGN can be 
represented by a power law with roughly equal
energy per decade from $10^{13}-10^{20}$~Hz (Elvis et al.\ 1994). 
This is much broader than 
the ensemble of spectra from stars, which is roughly the sum of 
blackbodies with an effective temperature of $10^3-10^5$~K. 
Superimposed on this power-law form are strong optical and UV 
lines from hydrogen, highly ionized C, N, and O, and a complex of 
low-ionization Fe lines. Thus, selecting sources on the basis of either 
their similarity to an AGN SED or their difference from a stellar SED 
is rather productive.

While the SED for optically-selected AGN is well studied, that for
radio, infrared (IR), and X-ray--selected sources is much less well 
documented. Many of the radio-selected and hard X-ray--selected AGN 
show no indications of non-stellar colors in ground-based UV/optical/IR
observations, and, thus, the use of the SED in these bands as an
AGN indicator will not find the objects.

\subsection{Optical Color Selection and the Presence of a Semi-stellar Nucleus}

\begin{figure}[t]
\centerline{\includegraphics[scale=0.5]{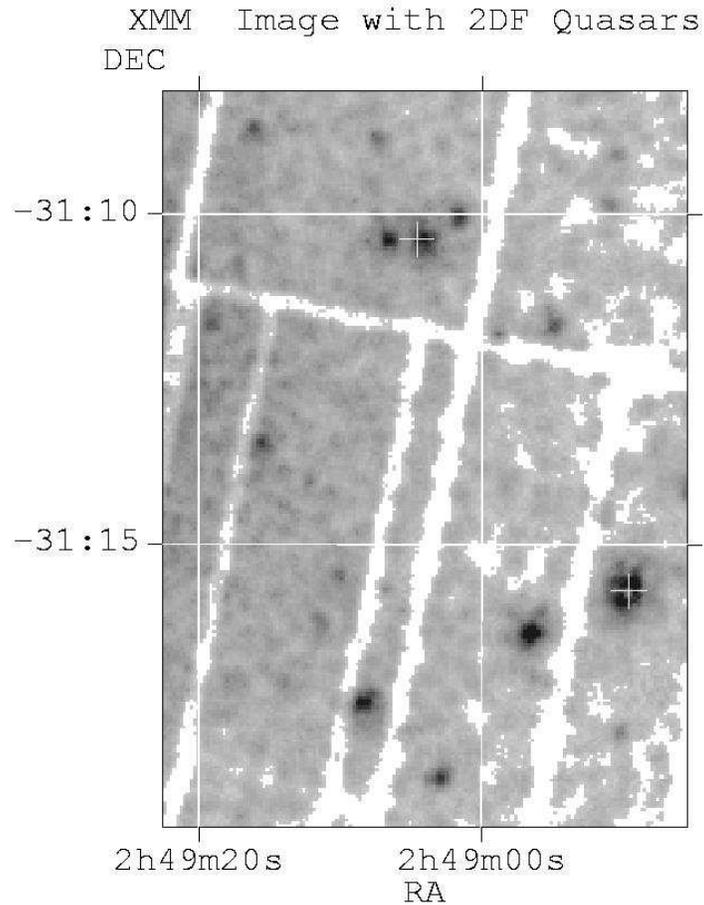}}
\caption{
{\em XMM-Newton\/} image of a 2dF quasar field with the 2dF
quasars marked by crosses. Notice the large numbers of X-ray point
sources, virtually all AGN, that are not identified in the 2dF.
}
\label{richardfig1}
\end{figure}

It was originally realized by Sandage (1971) that the optical colors
of an AGN (modeled as a power law plus broad emission lines) and the
starlight from a host galaxy deviated in a systematic way from pure
stellar colors as a function of power-law index, redshift, and the
fraction of total light in the non-thermal continuum. Using this
technique and multicolor data, it is possible to construct color
ratios that efficiently select objects with unusual colors (e.g., Koo
\& Kron 1988). These are then candidates for follow-up optical
spectroscopy and radio and X-ray observations. This technique has
reached its present apogee with the SDSS, which finds objects by 
their deviations in five-color space (Fan 1999; Richards et al.\ 2001). 
The technique is essentially ``trained'' on known objects,
and the colors of the quasars so obtained are sufficiently narrowly 
distributed that photometric redshifts can be estimated. While there 
are claims of ``completely unbiased'' optical samples (Meyer et al.\ 2001),
this reviewer believes that they are rather overstated. In particular,
X-ray images with the {\em XMM-Newton\/} and {\em Chandra X-ray
Observatories\/} of the 2dF fields find many AGN not detected by 
this optical survey (see Fig.~\ref{richardfig1}).

%
%

Over the last 30 years, there have been a large number of programs
(see Hartwick \& Schade 1990) that selected AGN using Schmidt's
(1969) criteria for selecting quasars:

\begin{itemize}

\item
Sources that do not have the colors of normal stars in 
their nuclei at any redshift. These AGN are found by 
``exclusion''.

\item
Sources that have a luminous semi-stellar nucleus 
(see Weedman 1977 for a description of a Seyfert~1 galaxy, 
and Sarajedini et al.\ 1999 for recent results).

\item
Sources that show time variability. It is well-known that
virtually all AGN vary (Schmidt 1969; Cannon, Penston, \& Brett 1971).

\item
Sources that show strong, relatively broad
UV/optical emission lines.

\item
Stellar sources that lack proper motion. This criterion has 
recently been revived by analyses of deep {\em Hubble Space 
Telescope (HST)\/} images.

\item
Luminous high-redshift sources that are selected by their 
optical colors via the Lyman break.

\end{itemize}

\noindent
These criteria are optimized for quasar-like sources and will not
select Seyfert~2-like sources very well. (Seyfert~2 galaxies have 
lines that are not as strong or as broad, nuclei that are not as bright, 
and they show little or no time variability in the optical bands.) 
The realization that AGN bolometric luminosity may not be well 
correlated with galaxy luminosity (Urry 2003) indicates that 
low-luminosity AGN are very difficult to find in massive galaxies 
via optical selection techniques. Essentially, the starlight dilutes 
the signatures of the AGN, reducing the equivalent widths of all of
the lines; at moderate redshifts ($z>0.2$), the starlight reduces 
or eliminates the color signatures of the nucleus.

Hartwick \& Schade (1990) review the results of many of these
surveys and provide an excellent summary of optical AGN
selection techniques.  However, they do not compare the results to
techniques obtained via other methods. It is somewhat amusing to find
stated in the Hartwick \& Schade (1990) paper that the samples found
by these optical techniques are a complete sample of quasars. As we
shall see below, optical techniques miss most radio-selected AGN
and a large fraction of IR and X-ray--selected AGN.

The presence of a semi-stellar nucleus was one of the original
defining criteria for quasars. However, this criterion is highly
sensitive to the angular resolution of the telescope and to the relative
contrast of the nucleus and the host galaxy. It also finds
bright super star clusters and other compact objects. Sarajedini et
al.\ (1999) searched for stellar nuclei in a relatively large
sample of galaxies serendipitously observed by {\em HST\/} and 
found that $\sim10$\% show unresolved semi-stellar nuclei that
contribute more than 5\% of the total optical light. So far, this 
is the best optical survey for low-luminosity AGN-like activity at 
moderate redshifts.

While optical techniques are very powerful, they have several
problems. Their use requires that the nuclei either be bright enough
to outshine the stars in the photometric extraction region or have
sufficiently different colors to be recognizable. The effects of
copious star formation on the optical colors are considerable. For
example, in the Byurakan surveys, which selected objects by searching
for blue continua, 90\% of the objects found are starbursts rather
than Seyferts. Thus, this selection technique is only five times more
efficient than a blind survey. The effect of stellar dilution has
been quantified for the SDSS by Richards et al.\ (2001), who noticed
that more luminous objects are bluer at a fixed redshift due
presumably to the increased importance of starlight in lower
luminosity objects. At an absolute magnitude in the Sloan $g$ band
of $-23$ (or an optical
luminosity of $6\times10^{44}$~ergs~s$^{-1}$), this effect seems to 
go away, indicating that the effects of starlight are minimal for these
extremely luminous objects. Thus, there are strong selection effects
with the luminosities of the nuclei and of the host galaxies, and with 
redshift. In principle, these selection effects are quantifiable, 
but in practice, this only works at low redshifts, where one can properly 
model the host galaxy and remove the starlight 
(Moran, Filippenko, \& Chornock 2002; see also Moran, this volume). 
Because there is no simple relation between the host galaxy and 
the nucleus properties, the problem is rather intractable. There are 
also large color selection terms because of the varying contrast 
between a ``typical'' stellar spectrum and the SED of an AGN with
redshift.

Color selection is a very efficient technique, and with a large database
like the SDSS, produces copious samples of AGN over a wide redshift
range. However, such methods must be evaluated carefully because of
the large ratio of stars to quasars. At $m\sim18$ there are $\sim500$
stars to every quasar, and thus the classification accuracy must be 
better than 0.2\% to avoid severe contamination! Therefore, techniques 
have focused on stellar ``rejection'' rather than on
``finding all the quasars'', unless extensive spectroscopic follow-up 
is also available for the survey.

In addition, since it is now well-known (see discussion in 
\S\ref{richardsecselection}) that 
many (most?) AGN have large amounts of dust and gas in the 
line of sight, the effects of extinction can be very large and have
to be very well modeled to produce reliable samples. The effects of
extinction can be partially ameliorated via the use of near-IR
data, but only up to some limit. It has been known for very many years 
that the reddening curve in AGN is different from that in the 
interstellar medium of the Milky Way (Maiolino et al.\ 2001), 
and thus it is not at all clear how to ``de-redden'' the AGN spectrum 
to derive reliable optical fluxes.

\subsection{Optical Emission Lines, Variability, and Zero Proper
Motion Selection}

There have been a wide variety of AGN surveys relying on 
low-resolution optical spectra. These techniques select AGN on the
strength of the UV/optical emission lines (see Osmer \& Hewett 1991
for a summary of this technique, and Salzer et al.\ 2000 
for the latest application with the Kitt Peak National Observatory
International Spectroscopic Survey). 
While this method essentially
has no ``false'' detections, its completeness is very difficult to
evaluate, since the signal-to-noise depends on the equivalent widths 
of the lines. Since the lines observed in the optical depend on the 
redshifts of the objects (e.g., from H$\alpha$ to Ly$\alpha$) and 
vary in intrinsic strength by over an order of magnitude, it is very
difficult to evaluate the completeness of the sample. Furthermore, the
effects of redshifting make the effective band smaller by $(1+z)$, such
that at $z=3$, only 1400~\AA\ in the rest-frame are covered by the total
ground-based optical wavelength band ($3200-8900$~\AA). The limit of 
this technique (Ho, Filippenko, \& Sargent 1995) is to obtain high 
signal-to-noise spectra of the nucleus of every object in a complete 
sample. However, even this method runs into the difficulty of removing 
absorption lines due to stars in the relatively low-luminosity 
AGN spectra and requires extreme care.

Variability can also be used as a survey technique
(Usher \& Mitchell 1978; Brunzendorf \& Meusinger 2002; 
Dobrzycki et al.\ 2003), since on long enough timescales, 
virtually all AGN are variable 
(Veron \& Hawkins 1995; Giveon et al.\ 1999). This method requires a 
large number of observations and seems to produce a lower areal 
density of objects than other techniques. However, this method also 
finds large numbers of variable stars and supernovae and must be 
combined with other selection criteria.

Similarly, since AGN are at great distances compared to bright stars,
the absence of proper motion can be a guide to AGN selection. However,
this technique has not been widely used and fails at faint magnitudes,
where most of the stars also do not show proper motion. This
technique will be resurrected when large solid angle astrometric 
data sets are available.

\subsection{Radio Selection}

Radio selection was the original way to find AGN. It is very
powerful, since almost all luminous radio sources are AGN in our
definition of radiating supermassive black holes---the only
contamination is at lower luminosities, where there can be
significant radio emission from rapidly star-forming regions. Thus,
for most luminous objects, mere detection in the radio indicates the
presence of an AGN. Radio surveys are also rather sensitive, and the
positions are extremely accurate. The morphological information is
also important, since all radio sources showing double-lobe or
jet-like structures are AGN.

However, radio surveys are very incomplete, since less than 10\% of 
AGN have luminous radio emission (White et al.\ 2000), the ratio of 
bolometric luminosity to radio luminosity varies
by a factor of $10^5$, and the fraction of the total energy radiated in
the radio is small. While it is true that most ``radio-quiet'' AGN
have some radio emission, the relative intensity is very low, making
searches for these objects in the radio very difficult. Even the
early radio samples found objects with a wide range of optical
properties and a wide redshift distribution. The absence of a 
correlation of radio flux with redshift indicated a broad luminosity 
function.

It was noted in the 1960s that a significant
fraction of the radio sources did not have broad lines or had no
lines at all and that their optical colors were not unusual
(Kristian, Sandage, \& Katem 1974). To quote
from Ulrich (1971), ``Comparison of the optical and radio properties
of nuclei of galaxies shows that galaxies with compact nuclear radio
sources are more likely to have optical spectra with emission lines
than are galaxies without central radio sources'', thereby indicating 
the relative rareness of broad or strong lines in radio-selected AGN.
This has been quantified recently in a combined study of the 2dF
survey with a southern radio survey (Sadler et al.\ 2002;
Magliocchetti et al.\ 2002), as well as in SDSS and FIRST
(Faint Images of the Radio Sky at Twenty-cm) data (Ivezic et al.\ 2002). 
In these data, over 60\% of nearby radio galaxies show no
evidence for strong emission lines of any width, and their optical
continua show weak or no evidence for non-thermal activity. 
This is also true at higher redshifts, where the radio sources
are not even found to be X-ray sources (Cowie et al.\ 2004a).
In the SDSS data, only 10\% of the optical counterparts of radio sources 
show evidence for strong AGN-like emission lines. However, in several of
the nearby objects for which no obvious nucleus or broad lines exist in
ground-based data, {\em HST\/} imaging finds evidence 
for weak non-thermal continua (Chiaberge et al.\ 2003). In fact, 
the absence of strong lines in the radio galaxies made optical
spectral identification so hard that it required more than 30 years 
to completely identify the first radio catalog 
(Spinrad et al.\ 1985).
It is quite interesting that the vast majority of radio
AGN would never be selected by optical techniques. This was the
first indication that different selection techniques tend to find
different objects. To rephrase this, while only 10\% of 
optically-selected AGN are ``radio loud'', less than 30\% of 
radio-selected AGN are ``optically loud''.

The ``radio'' colors of AGN allow discrimination against other sources
of radio emission for luminous objects.  Almost all flat-spectrum,
compact radio sources are AGN. In addition, the strong correlation of
mid-IR with radio flux found for star-forming galaxies allows 
the detection of AGN by exclusion, if one also has IR data 
(Yun, Reddy, \& Condon 2001). That is, the objects with low 
IR-to-radio luminosity ratios are almost all
AGN (Condon, Anderson, \& Broderick 1995), and those objects that 
lie in a narrow range of high IR-to-radio luminosity ratios are 
starbursts (or radio-quiet AGN). This technique has been 
used to show that many of the objects thought to be starbursts in the 
{\em ROSAT\/} All-Sky Survey data were really previously uncataloged 
AGN (Condon et al.\ 1998).  However, inspection of a large sample of 
X-ray--selected objects (the most famous of which is NGC1068) shows 
many of them with ``star formation'' IR-to-radio ratios, indicating 
that most of the IR and radio luminosities are not AGN related. 
It thus requires a good deal of care (Miller \& Owen 2001) to select 
the AGN, which tend to have lower relative far-IR--to--radio 
luminosities and larger scatter in the correlation than those seen for 
star-forming galaxies.

In an unbiased survey of radio-selected objects using the SDSS 
(Ivezic et al.\ 2002), the number
of starburst galaxies outnumbers those objects with
AGN-like optical spectra by 5:2, and the number of objects without
any unusual optical properties dominate the optically-active AGN by
12:1. It thus seems that deep radio surveys have a large
contamination factor from non-AGN populations but discover lots of
AGN that would not be selected with classical optical criteria.

Most radio data have fairly accurate positions (better than $45''$ for
the largest radio survey, the NRAO VLA Sky Survey or NVSS, and better 
than $7''$ for the most sensitive large solid angle survey, FIRST), 
allowing counterparts in other wavelength bands to be readily 
identified. However, only $\sim8$\% of optically-selected AGN 
(Ivezic et al.\ 2002) are radio ``loud'' and selectable in radio 
surveys.

\subsection{Infrared Selection}

The use of IR techniques to measure AGN continua started in the
1970s with the advent of the first sensitive IR detectors (Low \&
Kleinmann 1968). However, the IR ``colors'' of Seyfert galaxies 
are only subtly different than those of normal galaxies (Kuraszkiewicz et
al.\ 2003), and the equivalent widths of the IR lines are not 
sufficient to use as a finding mechanism. Thus, IR color surveys 
can have a large fraction of ``false'' AGN, unless great care is 
taken.

The first large-scale attempt to find AGN in the IR was based 
on {\em Infrared Astronomical Satellite (IRAS)\/} data. 
De Grijp, Lub, \& Miley (1987) 
showed that AGN had systematically different $60\mu$m/$25\mu$m 
colors than normal galaxies. An alternative approach
(Spinoglio \& Malkan 1989) is to obtain optical spectra of every 
IR-selected galaxy. This is a follow-up of the idea of 
Huchra \& Burg (1992) to obtain optical spectra of every 
optically-selected galaxy, but it is not really a survey technique.  
The latest use of the IR to find active galaxies is with 
the Two Micron All Sky Survey (2MASS; Cutri et al.\ 2002).  
In this survey, $\sim60$\% of the 
objects with $J-K>2$ are found to have the optical properties of AGN. 
This selection criterion is bootstrapped by using the near-IR 
colors of known radio and optically-selected AGN (Elvis et al.\ 1994),
and thus will tend to find objects with similar properties. The large 
space density of these IR-selected objects makes them a major 
contributor to the AGN population.

One of the main potential advantages of the IR is that, at 
least for known AGN, the mid-IR ($3-10~\mu$m) seems to be a
``pivot point'' in the SED. That is, objects of the same bolometric 
luminosity have rather similar mid-IR luminosities but very 
variable radio, optical, UV, and soft X-ray luminosities. 
Exactly how to use this information in a survey via
color-color plots has been investigated by Andreani et al.\ (2003)
and potentially seems to be quite powerful. However, to avoid large
numbers of false detections, IR selection requires very 
broadband, precision IR photometry from $4-100~\mu$m.
Because of the large ratio of normal to ``active'' galaxies, 
there can be ``leakage'' in the selection criteria, unless 
photometric errors are small.

It has been demonstrated (de Grijp et al.\ 1987) that most
optically-selected AGN have a ``mid-IR excess'' and are ``warmer''
compared to normal galaxies. This also seems to be true for 
radio-selected AGN. 
The effect is very subtle, however, with AGN tending to
have a ``hotter'' mid-IR effective dust temperature. However, 
based on {\em Infrared Space Observatory (ISO)\/} colors of AGN 
(Haas et al.\ 2003), it is not clear how complete this technique is, 
since the mid-IR slopes span a range from $0.9-2.2$, and
the relative brightness of the mid-IR AGN component to the 
stellar continuum and star-heated dust spans a wide range. At long 
IR wavelengths, the contribution from normal stars fades, and 
there is emission from AGN and star formation-heated dust.

There is a long-running discussion in the IR community about 
how to separate out dusty AGN from starburst galaxies based on 
optical and IR data (see Veilleux 2002 for an excellent summary 
of the present situation). In this reviewer's opinion, it is 
basically impossible without IR spectroscopy. In the mid-IR,
the presence and absence of IR emission features associated
with polycyclic aromatic hydrocarbons (PAHs) and high-ionization 
IR lines (Laurent et al.\ 2000) provide strong AGN diagnostics.  
However, there have not been any wide-angle IR spectroscopic 
surveys designed to find active galaxies. This will change with the
{\em Spitzer Space Telescope\/}. Because of a lack of data, it is 
not known if there are mid-IR ``quiet'' AGN.

Infrared observations of X-ray--selected (Kuraszkiewicz et al.\ 2003) 
and optically-selected (Haas et al.\ 2003) AGN samples show a very wide 
range of IR to X-ray SEDs and considerable variation in the 
IR colors of AGN, making the use of IR colors somewhat 
problematic in an AGN survey. The IR spectral parameters do 
not correlate with the optical spectral slope, nor with the IR 
luminosity, nor with the mid-IR--to--far-IR luminosity 
ratio, nor with inclination-dependent extinction effects in the 
picture of a dusty torus.

\subsection{High-Energy Selection}

The other two techniques used to find active galaxies are X-ray and
$\gamma$-ray--selection. Like radio emission, luminous, compact X-ray and
$\gamma$-ray emission is an almost certain indicator of the existence of an
AGN and does not need ``confirmation'' by data in other wavelength
bands. The relatively low signal in most X-ray surveys makes X-ray
spectroscopic redshift determinations difficult, and for most objects
one must rely on optical redshifts. However, in a few recent cases,
X-ray redshifts have been determined (Hasinger 2004).
The need to obtain optical redshifts has driven the entire field of 
``optical identification'', in a fashion similar to that of the 
identification of radio surveys.

It was realized in the early 1970s (Pounds et al.\ 1975) that 
about one-quarter of the high-latitude X-ray--selected objects 
in the {\em Ariel-V\/} X-ray survey could be identified with previously 
known AGN. Given the large
uncertainties in the positions of these objects ($\sim0.5-2$~deg$^2$),
this was quite surprising. Detailed follow-up work of the unidentified, 
high-latitude X-ray sources in the early surveys 
(Ward et al.\ 1980; Wilson 1979) 
discovered that most of them were previously
unknown AGN with properties that ``hid'' them from optical surveys.
This work was strengthened with the first accurate X-ray positions
(Griffiths et al.\ 1979), which confirmed that objects with rather weak
or narrow optical lines and no evidence for a non-thermal continuum
could be luminous X-ray sources. The advent of X-ray imaging with the
{\em Einstein Observatory\/} in 1979 vastly increased the efficiency
and accuracy of X-ray surveys, but, with the exception of 
low-redshift objects, the positions were still not accurate enough to
have an unique optical counterpart. This drove a very large program of
identification (the {\em Einstein\/} Medium-Sensitivity Survey or EMSS; 
Gioia et al.\ 1990), which took several years to complete.

A major difference in the early X-ray and optical surveys was the
redshift distribution, with optical selection criteria finding many
objects over a wide redshift range out to $z\sim3$, while X-ray samples
were much more concentrated at $z<1$. Even the early X-ray samples
found very little correlation of redshift with flux.

So far, while $\gamma$-ray emission is certainly an almost unique feature 
of high-latitude active galaxies, the positions are too poor to allow
the identification of the objects. Based on correlations with radio data,
most of the identified AGN are blazars or flat-spectrum radio sources,
and very few, if any, classical Seyfert~1 galaxies or quasars are found 
in the {\em Compton Gamma-Ray Observatory\/} surveys.

It has been known for over 35 years that the vast majority of 
high-latitude, point-like ``hard'' X-ray sources are
AGN\footnote{In a historical point, before the {\em Ariel-V\/}
results, there were strong indications from {\em Uhuru\/} data of 
a class of ``unidentified high galactic latitude X-ray sources'' 
(Holt et al.\ 1974). Arguments---later proved false---were put forward  
that they could not be Seyfert galaxies.} (Pounds 1979), and
this is one of the most efficient and least ``error prone'' ways of
selecting AGN. At the present time, there are very few, if any, known
AGN that are not also luminous X-ray sources. However, because of
the wide variance in the SEDs of AGN, there are many objects which 
do not have sufficiently sensitive X-ray data to decide if they
really are X-ray ``quiet'' objects (Leighly, Halpern, \&
Jenkins 2002; Gallagher et al.\ 2001). 
There are classes of AGN (in particular, broad absorption-line
quasars and strong FeII objects; Lawrence et al.\ 1997) that 
have rather low soft X-ray--to--optical ratios. At present, this 
is thought to be due to the high column densities in these objects 
(see \S\ref{richardsecselection}), but work is still proceeding 
on this. Recent results suggest that there is a class of IR-selected 
AGN with broad optical emission lines that are X-ray ``weak'' 
(Wilkes et al.\ 2002).

The efficiency of X-ray surveys is very high, finding considerably
more AGN at a fixed optical magnitude than other techniques. The
asymptotic limit of optical surveys is $\sim130$~AGN~deg$^{-2}$ 
at $B<23$~mag (Palunas et al.\ 2000), while the {\em Chandra\/} 
surveys find $\sim1000$~deg$^{-2}$ at $r<24$~mag.  
This relative efficiency is true even for very optically-bright 
samples; for example, Grazian et al.\ (2000) finds three
times more AGN with $v<14.5$ in the {\em ROSAT\/} All-Sky Survey 
than does the color-selected Palomar-Green (PG) bright quasar survey 
(Schmidt \& Green 1983). In the $2-8$~keV band at fluxes between
$10^{-15}$ to $10^{-10}$~ergs~cm$^{-2}$~s$^{-1}$, essentially all 
of the point-like sources are AGN. At lower fluxes, the
fraction of star-forming galaxies increases, but the X-ray counts 
are still dominated by AGN over the range to which {\em Chandra\/} 
has so far reached.

It is only recently with {\em Chandra\/} and {\em XMM-Newton\/} data 
that the X-ray/optical/near-IR distribution of normal galaxies 
has been measured, and thus an estimate of the AGN ``contamination 
factor'' derived. The bottom line is that all compact X-ray sources 
with a luminosity above $\sim10^{42}$~ergs~s$^{-1}$ ($2-10$~keV)  
are considered to be AGN. The main interlopers are 
``ultraluminous X-ray sources'' (ULXs; Colbert \& Ptak 2002), which 
are objects with X-ray luminosities between $4\times10^{39}$ 
and $10^{42}$~ergs~s$^{-1}$ and do not reside in the nuclei of 
the host galaxies. While easily recognized at low redshifts,
{\em Chandra\/} imaging is required to separate these out at $z>0.1$ 
(Hornschemeier et al.\ 2003). At X-ray luminosities below 
$10^{42}$~ergs~s$^{-1}$, there is always the possibility that a
nuclear source might really be an ULX rather than an AGN, as is
apparently the case in M33 (Long, Charles, \& Dubus 2002).
However, it is not clear
if the distinction is meaningful if the ULXs turn out to be 
supermassive black holes. Starburst galaxies have a considerably 
lower X-ray--to--optical flux ratio than AGN, and while they 
frequently harbor ULXs (Grimm, Gilfanov, \& Sunyaev 2003), their 
total luminosities rarely reach $10^{42}$~ergs~s$^{-1}$ and are 
well correlated with their IR luminosities (Ranalli, Comastri,
\& Setti 2003).

For ``typical'' soft X-ray--selected AGN, the X-ray--to--$B$-band 
flux ratio is $\sim1$, with a full range of $\sim100$ 
and a variance of $\sim6$ (Mushotzky \& Wandel 1989;
Anderson et al.\ 2003). In the $2-10$~keV X-ray band, there is a 
broader range of X-ray--to--optical flux ratios, with a significant 
tail of very high X-ray--to--optical ratios that extends out to
$>10^4$ (Akiyama et al.\ 2003; Barger et al.\ 2003b; 
Comastri et al.\ 2003; see also Comastri, this volume).

Because of the broad wavelength coverage of the X-ray band (about a
factor of 100 in wavelength), there is a significant difference between
soft-band ($0.2-2$~keV), hard-band ($2-6$~keV), and very hard-band 
($5-10$~keV) surveys. The soft-band surveys can suffer from the same 
extinction effects as UV and optical surveys, while in the harder bands,
absorption is a much smaller effect. Based on {\em ROSAT\/} and 
{\em Chandra\/} data, the soft and hard X-ray selection tends to find 
different objects. The soft X-ray selection preferentially finds 
broad-line quasars and narrow-line Seyfert galaxies (these objects
tend to have a bright, soft spectral component, in addition 
to a flat, high-energy, power-law spectrum).
The hard X-ray selection finds, in addition to the classical Seyfert~1
galaxies and quasars, large numbers of objects with weak or absent 
optical emission lines and lacking non-thermal nuclei.
It is believed that this selection effect may be related to the presence 
of absorbing material, but other possibilities exist 
(see \S\ref{richardsecselection}). Large
follow-up programs for X-ray--selected AGN find other selection
effects. The median value of the soft X-ray--to--optical flux ratio changes 
by a factor of $2-3$ over four orders of magnitude in optical flux 
(Anderson et al.\ 2003). Thus, high-luminosity objects are somewhat 
weaker in the X-ray band for a given optical luminosity. On the other hand,
radio-loud AGN have a factor of three higher X-ray--to--optical flux ratio 
than radio-quiet objects (Worrall 1987). Thus, soft X-ray selection compared 
to optical selection is biased towards radio-loud, lower luminosity objects.
While there is a fair range in the X-ray spectral slopes of AGN (the
variance is $\Delta \alpha\sim0.2$), there is no correlation of slope 
with luminosity or redshift (Vignali et al.\ 2003; Reeves et al.\ 1997).

Before the {\em Chandra\/} and {\em XMM-Newton\/} surveys, X-ray error 
circles were often large (e.g., $>15''$), engendering large optical 
follow-up programs and allowing a bias in the selection of the 
optical ``counterpart''. However, {\em Chandra\/} and {\em XMM-Newton\/} 
follow-up of {\em ROSAT\/} identifications (McHardy et al.\ 2003) show 
that virtually all of them were ``right'', and, thus, this possible 
selection effect is small.

\subsection{Ultraviolet Selection}

While it has long been possible to select high-redshift objects in
the rest-frame UV---and this is often the main path to ``optical''
selection of quasars---until 2003 there has been no large-scale survey
in the UV for active galaxies. This will be remedied with the
recent successful launch of the {\em Galaxy Evolution Explorer (GALEX)\/} 
satellite. To date, there has not been a rest-frame UV survey 
for AGN covering a reasonable solid angle with reasonable sensitivity.

\section[Selection effects]{Selection Effects}
\label{richardsecselection}

The whole subject of finding and cataloging AGN is dominated by
selection effects (see Hewett \& Foltz 1994 for an extensive discussion 
and an excellent summary of the field). The main effects are: 

\begin{itemize}

\item[(1)] 
{\em Dilution of the optical/IR brightness and color by starlight.\/} 
This is a large effect for AGN with optical/IR luminosities less 
than that of an average galaxy (roughly $10^{44}$~ergs~s$^{-1}$) at 
moderate to high redshifts. This effect has made optical selection 
of Seyfert galaxies at $z>0.5$ a very difficult process, since it 
reduces the signal from time variability, color, or emission-line 
survey techniques. This effect is unimportant for X-ray selection 
above a luminosity of $\sim10^{41.5}$~ergs~s$^{-1}$ and for radio 
selection above a power of $10^{24}$~Watts~Hz$^{-1}$. The dilution 
reduces the equivalent widths of all of the lines, as
well as the visibility of the non-thermal radiation in the spectral
bands where stars are luminous (i.e., optical/near-IR). The magnitude of
this effect clearly depends on angular resolution and aperture and is
minimized for {\em HST\/} data. 
However, very recent work shows that the
brightness of the optical nuclei for the hard X-ray--selected sources
in the {\em Chandra\/} Deep Field-South are a factor of 20 dimmer than
expected based on the {\em ROSAT\/} sources (Grogin et al.\ 2003). 
It may also be important for luminous AGN at high redshifts, where 
most galaxies seem to have copious star formation.\footnote{Recent work
by Comastri et al.\ (2003) shows that X-ray--to--optical luminosity 
ratios for objects having only narrow optical lines rise as 
$L_{2-10~{\rm keV}}$).
Note that these may not be the same as classical Seyfert~2 galaxies; 
for the purposes of this chapter, I follow the nomenclature of the 
{\em BeppoSAX\/} High-Energy Large Area Survey (HELLAS) team
and call these optically-obscured type~2 AGN.
Thus, at high X-ray luminosities, these objects are X-ray bright
rather than X-ray dim.}

\item[(2)] {\em Obscuration.\/}
Based on X-ray and IR surveys, one finds that many (most?)
AGN have large column densities of gas and dust in
the line of sight (Fabian \& Iwasawa 1999).
Models of the X-ray and IR backgrounds (Almaini, Lawrence, \& Boyle 1999) 
suggest that more than 70\% of all AGN have high column
densities ($>10^{22}$~atoms~cm$^{-2}$) in the line of sight, which,
for normal dust-to-gas ratios, gives an effective optical absorption
of $A_V>5$, effectively extinguishing the UV/near-IR fluxes. However, 
the situation is not simple: there are many objects known with 
(a) high X-ray column densities and 
luminous UV continua (the most famous being NGC4151), 
(b) high IR dust emission and low X-ray column densities 
(e.g., IRAS~1334; Brandt et al.\ 1997), and (c) very red continua, 
strong, broad optical lines, and apparently very high X-ray column 
densities (e.g., Wilkes et al.\ 2002; for a review, 
see Comastri et al.\ 2003). 
The multivariate distribution functions of each of these is not known 
at the present time, and thus these effects cannot be corrected for.

In unified models of active galaxies (Antonucci 1993), the physical
difference between Seyfert~1 and Seyfert~2 galaxies is that the 
line of sight to the latter is blocked by optically-thick material 
in the UV and optical. This accounts for the observed weakness
of UV/optical/soft X-ray emission, the lack of short timescale 
intensity variability, the high optical polarization, and the detailed 
form of the X-ray spectra. For these objects, the observed 
UV/soft X-ray continuum is only a small fraction of the emitted 
radiation, and much of the energy is emitted in the IR. 
Seyfert~2 galaxies have low soft X-ray luminosities, low optical 
luminosities, and large IR luminosities. Thus, they are not found in 
great numbers in soft X-ray or optical color samples.

\end{itemize}

Both of the above selection biases make it very difficult to directly
connect samples derived in different spectral bands and to derive
unique values for the bolometric luminosity function and its
evolution. It is also not clear if these selection effects are
functions of redshift and source luminosity.

Much recent progress has been achieved via {\em Chandra\/} and 
{\em XMM-Newton\/} observations of optically well-surveyed fields, 
with some surprising results. The {\em Chandra\/}-selected AGN sample 
shows that at high X-ray luminosities, almost all AGN show broad 
optical emission lines, indicating that the effects of obscuration 
are small, while at low luminosities, the vast majority of AGN show 
little or no activity in the optical (Steffen et al.\ 2003). There 
are also strong indications of evolution in these ratios.

There is another important effect, which, while not a classical
selection effect, has important consequences for the nature and
completeness of AGN samples. It is clear that the
strengths and widths of various UV and optical lines are not 
random but lie along two eigenvectors (Boroson 2002). 
These line strengths are strongly correlated with the slope of 
the X-ray continuum (Brandt \& Boller 1998). Thus, in X-ray 
flux-limited surveys at different energy ranges, or in UV/optical 
surveys that rely on line widths and fluxes, one will end up with 
different sets of objects. There also may be evolution in the nature 
of these eigenvectors. It is speculated that the narrow-line Seyfert~1 
galaxies that lie along one of the extrema are radiating near the 
Eddington limit, and, if so, should be more common at higher redshifts.

\section[X-ray Selection of AGN]{X-ray Selection of AGN}

The advantages of X-ray selection of AGN include

\begin{figure}[t]
\centerline{\includegraphics[scale=0.45]{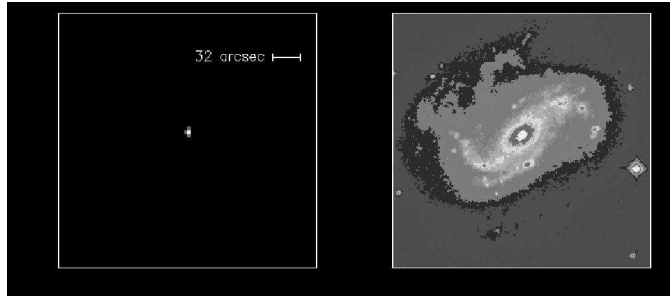}}
\caption{
{\em (Left panel)} X-ray ({\em ROSAT\/} High-Resolution Imager) 
and {(\em Right panel)} optical (SDSS) images of the nearby Seyfert
galaxy NGC4051. Notice in the X-ray image that essentially the only
observed emission is due to the AGN, while the optical image is
dominated by starlight. This is one of the original objects identified
by Seyfert (1943).
}
\label{richardfig2}
\end{figure}

\begin{figure}[t]
\centerline{\includegraphics[scale=0.5,angle=-90]{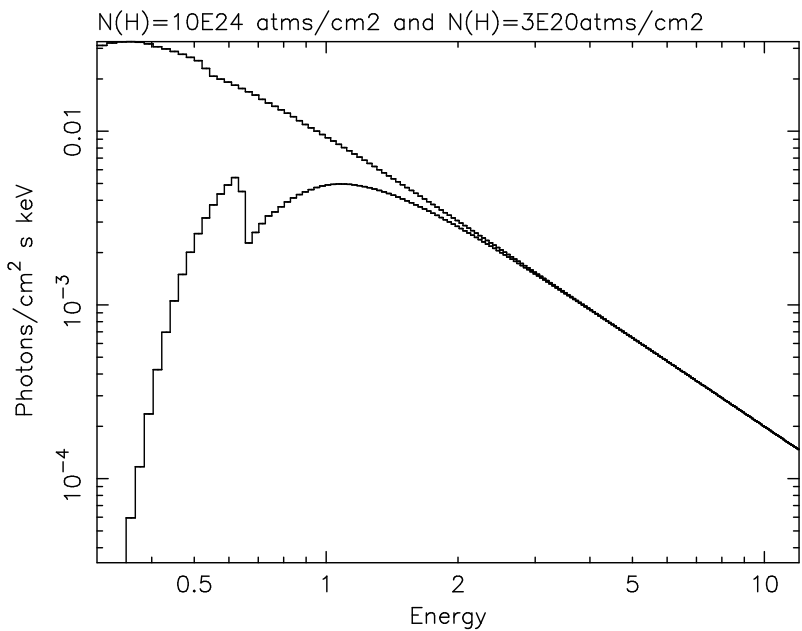}}
\caption{
X-ray spectra of two AGN at $z\sim10$, one with no
absorption, and the other with a line of sight column density 
of $\sim10^{24}$~atoms~cm$^{-2}$ and pure photoelectric
absorption. Note that at energies greater than
2~keV (observer's frame), there is no reduction in the source flux.
The effects at lower energies are much larger.
}
\label{richardfig3}
\end{figure}

\begin{itemize}

\item 
High contrast between the AGN and the stellar light
(see Figure~\ref{richardfig2}).

\item
Penetrating power of X-rays. Even column densities
of $3\times10^{23}$~cm$^{-2}$ (corresponding to $A_V\sim150$~mags) 
do not reduce the flux at $E>5$~keV significantly 
(see Figure~\ref{richardfig3}).

\item
Great sensitivity of {\em Chandra\/} and {\em XMM-Newton\/}. 
Sources in the luminosity range $10^{42}-10^{46}$~ergs~s$^{-1}$ 
can be detected out to $z\sim3$, independent of the nature of the 
host galaxy (e.g., Steffen et al.\ 2003).

\item
Accurate positions from {\em Chandra\/}. Unique identifications
can be made with counterparts in other wavelength bands. 

\item
A relatively large fraction of the bolometric energy ($3-20$\%)
is radiated in the classical X-ray bands (Ho 1999).

\item
High areal density of X-ray--selected AGN reaching
400~sources~deg$^{-2}$ at $F_X\sim10^{-15}$~ergs~cm$^{-2}$~s$^{-1}$
in the $2-8$~keV band (Moretti et al.\ 2002), a level easily reachable 
with {\em Chandra\/} and {\em XMM-Newton\/} in moderate exposures, 
compared to the maximal value of $\sim150$~deg$^{-2}$ in optical surveys.

\item
Large amplitude and frequency of variability in the
X-ray band (see Dobrzycki et al.\ 2003 for an interesting comparison 
of objects selected by X-ray imaging and optical variability 
techniques).

\end{itemize}

%
%

In contrast to the optical, where stellar light is a major
contributor, or the UV, where light from young massive stars 
often dominates, or the IR, where dust reradiation from 
massive stars dominates, or the radio, where emission from HII 
regions, young supernovae, and other indicators of rapid star 
formation are often very important, there are very few sources of 
radiation that can confuse the issue in the hard X-ray band.

Point-like X-ray emission is easy to recognize as being caused by 
low-luminosity AGN. Using surveys of the low-redshift 
universe as a guideline, if the total integrated X-ray luminosity of a 
small ($<2$~kpc in size) object is greater than $10^{42}$~ergs~s$^{-1}$, 
then the object is almost certainly an AGN. In the low-redshift universe, 
there are no galaxies with a total (non-AGN) luminosity exceeding 
this level. Thus, even without detailed X-ray spectra or imaging, 
the identification of the nature of the source is clear.

X-rays are also rather penetrating. Column densities
corresponding to $A_V=5$ ($N_H\sim10^{22}$~atoms~cm$^{-2}$) only 
reduce fluxes by $\sim3$ in the {\em Chandra\/} and 
{\em XMM-Newton\/} soft X-ray bands.
One can see in the $2-10$~keV X-ray surveys that
approximately half of the brightest objects are highly
reddened in the optical and often invisible in the UV.  
At $z\sim10$ the absorber has to be Compton-thick ($A_V\sim2000$!) 
to ``kill'' the X-ray flux (see Fig.~\ref{richardfig3}). 
Thus, there are no dark 
ages for very high-redshift AGN in the X-ray band caused by the 
Gunn-Peterson effect. X-rays have a ``reverse'' Ly$\alpha$ forest 
effect---redshifting reduces the effects of absorption. Thus, for 
a fixed flux and column density, high-redshift quasars are easier 
to detect. This effect is similar, but of smaller magnitude, 
to that seen for the submillimeter sources.

%
%

From a more physics oriented point of view (Mushotzky, Done, \&
Pounds 1993), the X-ray emission originates from very close to 
the central black 
hole, often shows large amplitude rapid variability, and is 
characterized by a non-thermal spectrum. Thus, the X-ray properties 
are directly connected to the black hole nature of the AGN and are 
not due to reprocessing of the radiation

The fundamental properties of black holes should not be functions of
metallicity or environment but only of mass, accretion rate, and
black hole spin. Since the X-ray flux originates from very close to
the event horizon, the X-ray properties of high-redshift ``primordial''
black holes should be very similar to that of lower redshift objects.
This allows a reasonable calculation of their observable properties
at high redshifts (Haiman \& Loeb 1999; see also Haiman \&
Quataert, this volume).

\subsection{Early X-ray Surveys}

The first large-scale surveys of the X-ray sky were performed by 
the {\em Uhuru\/} (Gursky \& Schwartz 1977) and {\em Ariel-V\/} 
(Pounds 1979) small satellites in the $2-6$~keV band in the 1970s. 
These surveyed the sky more or less uniformly above a flux threshold 
of $\sim2\times10^{-11}$~ergs~cm$^{-2}$~s$^{-1}$ and,
while non-imaging, were able to derive error boxes
small enough to identify many of the sources. One of the early
surprises was that approximately half of all the AGN identified by 
{\em Ariel-V\/} had not been identified previously
as active galaxies by radio or optical
surveys and had rather different properties (weaker non-thermal
continua, narrower weak lines, strong reddening) than 
optically-selected AGN. 
The last large solid angle survey in the $2-10$~keV band
was performed by the {\em HEAO-1\/} satellite, with the largest samples 
being from the A2 (Piccinotti et al.\ 1982) and A1 (Remillard et al.\ 1986)
experiments. These surveys produced a list of $<200$ AGN and a complete
identification of the Piccinotti et al.\ (1982) list, 
which had only 35 AGN.

\subsection{Soft X-ray Surveys}

\begin{figure}[t]
\centerline{\includegraphics[scale=0.5]{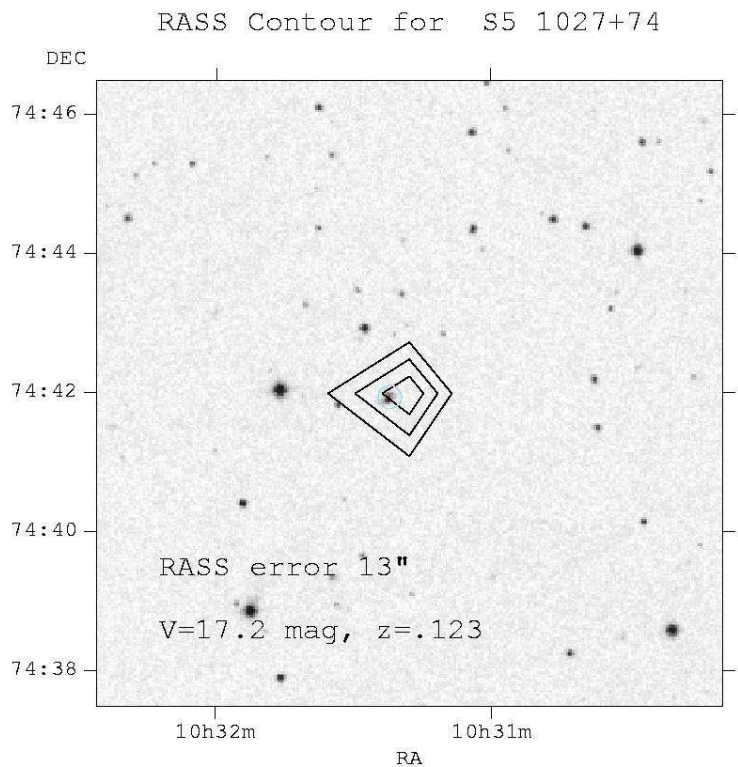}}
\caption{
X-ray contours from the {\em ROSAT\/} All-Sky Survey superimposed
on a field from the Palomar sky survey for the quasar S51027+74.
Note that the {\em ROSAT\/} contours are small enough for a 
high-probability optical identification of the source.
}
\label{richardfig4}
\end{figure}

The {\em Einstein\/} and, especially, the {\em ROSAT\/} surveys have 
provided very large samples of soft X-ray--selected AGN. There have
been extensive discussions of samples obtained by these missions (see
Puchnarewicz et al.\ 1996 for a summary of the pointed observations, 
and Fischer et al.\ 2000 and Zickgraf et al.\ 2003 for the survey data). 
At moderate X-ray fluxes ($>10^{-13}$~ergs~cm$^{-2}$~s$^{-1}$), there 
is only $\sim1$ X-ray source per square degree, and the {\em ROSAT\/} 
PSPC\footnote{The PSPC is the Position Sensitive Proportional Counter,
the workhorse instrument on {\em ROSAT\/} with a $15''$ spatial 
resolution and $3''-5''$ positional errors.} had sufficiently small 
positional errors that unique identifications could be made for the 
sources brighter than $m\sim20$~mag in the $B$ or $V$ bands
on the basis of the X-ray positions alone (see Fig.~\ref{richardfig4}). 
However, at lower fluxes, more accurate positions and
better angular resolution are often required, and many of the
identifications were made on the basis of finding a broad-line object
inside the X-ray error circle. Historically, this is how the previous
large soft X-ray survey with {\em Einstein\/} (the EMSS;
Gioia et al.\ 1990) obtained optical counterparts.
At faint X-ray fluxes ($<10^{-14}$~ergs~cm$^{-2}$~s$^{-1}$),
very long observations were required to detect the sources (the 
{\em ROSAT\/} Deep Surveys; Hasinger et al.\ 1998). 
It is historically interesting that the identifications
of most of the {\em ROSAT\/} Deep Survey sources were just at the limit of 
the capabilities of the largest ground-based optical telescopes, with 
the optical identification of the survey being almost complete at
$R\sim23$~mag (Hasinger et al.\ 1999).

%
%

\begin{figure}[t]
\centerline{\includegraphics[scale=0.5,angle=-90]{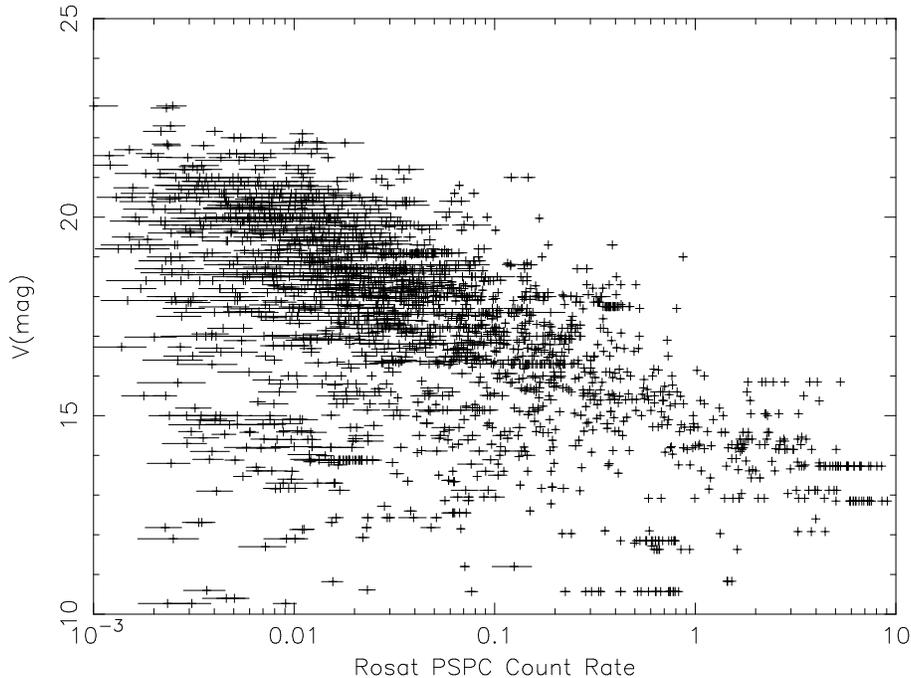}}
\caption{
Correlation of the {\em ROSAT\/} All-Sky Survey counting rate
and optical magnitude for a large sample of active galaxies. Note
that the majority of objects lie in a narrow band in magnitude
vs. X-ray counts space. 
}
\label{richardfig5}
\end{figure}

Soft X-ray surveys find that there is a moderate correlation of
optical and X-ray properties (see Fig.~\ref{richardfig5}) 
with a relatively narrow ($\pm1$ order of magnitude) range in 
X-ray--to--optical flux ratios,
that the X-ray evolution of AGN is similar to that of optically-selected 
AGN (see Miyaji, Hasinger, \& Schmidt 2000), and that most of the objects 
are broad-line AGN (see Hasinger et al.\ 1999 for a review). In the shallow 
{\em ROSAT\/} All-Sky Survey (Appenzeller et al.\ 2000), the median 
redshift is $z\sim 0.2$, and there are only two objects at $z>2$. 
This is to be contrasted with the ``shallow'' optical Bright Quasar
Survey (BQS; Schmidt \& Green 1983), which has a much flatter redshift 
distribution out to $z\sim 2$. However, there is a significant fraction 
of unusual AGN, some of which have rather red continuum colors 
and broad optical emission lines (Puchnarewicz \& Mason 1998).

%
%

As discussed in \S\ref{richardsecselection}, 
there are significant selection effects in soft 
X-ray surveys caused by obscuration in the line of sight. Perhaps the 
most direct example of this is the analysis of {\em ROSAT\/} 
observations of the Piccinotti et al.\ (1982) hard X-ray survey 
(Schartel et al.\ 1997), which shows that approximately half 
of all the hard X-ray sources are absorbed and thus have
significantly lower soft X-ray fluxes---up to 300 times less---than
expected if the spectra were not absorbed. A comparison of the hard
X-ray properties of a soft X-ray--selected sample has not been done
because of the limited sensitivities of large solid angle hard X-ray
surveys.

As one goes to higher redshifts, the effect of obscuration in the
X-ray band decreases as the spectrum redshifts (Barger et al.\ 2001),
so the observed $0.5-2$~keV band is much less affected by
obscuration effects. This seems to be a rather important effect for
the {\em Chandra\/} sources, and thus the {\em ROSAT\/} soft-band data 
need to be corrected for the changing effective bandpass.

{\em ROSAT\/} also performed a survey in the softest X-ray bands at
0.25~keV (Vaughan et al.\ 2001). Because of the effect of 
obscuration by the galactic column density in this low-energy 
range, only $\sim0.6$~sr could be covered. The sample appears to 
be complete. The identifications are rather different than in other 
AGN surveys, with one-third of the AGN identifications being BL Lacertae 
objects and one-quarter being narrow-line Seyfert~1 galaxies. These
are relatively large percentages for these two classes of objects
compared with surveys in any other wavelength band. Such a survey 
has a bias towards these objects because of their steep X-ray spectra. 
The almost disjoint nature of the objects contained in a hard (see
\S\ref{richardsecpre}) and extremely soft X-ray selection is a 
warning to the completeness of any survey.

\subsection{Pre-{\em Chandra} and {\em XMM-Newton} Hard X-ray Surveys}
\label{richardsecpre}

The most sensitive large solid angle $2-8$~keV surveys have been
obtained via serendipitous sources detected by {\em ASCA\/} 
(Akiyama et al.\ 2003; Nandra et al.\ 2003) and {\em BeppoSAX\/} 
(Fiore et al.\ 2000). These satellites had
moderate fields of view and moderate sensitivities,
reaching $2-10$~keV flux limits of
$\sim2\times10^{-13}$~ergs~cm$^{-2}$~s$^{-1}$. At this flux
level, there are $5-10$ sources per square degree 
(Cagnoni, della Ceca, \& Maccacaro 1998; Ueda et al.\ 2001), 
or roughly one serendipitous source per {\em ASCA\/} or
{\em BeppoSAX\/} pointing. Over its lifetime, {\em ASCA\/} 
obtained $\sim500$ serendipitous sources 
over $\sim100$~deg$^{2}$, and {\em BeppoSAX\/} somewhat fewer. 
Because of their moderate $\sim1-3'$ FWHM angular
resolutions and $40-100''$ angular accuracies (similar to those of the 
earlier {\em Einstein\/} soft X-ray serendipitous survey), 
optical follow-up has been tedious, and the number of identified sources 
is now only $\sim150$ objects. The difficulty of optical,
IR, and radio follow-up seriously delayed the results from these 
surveys, and they appeared after the initial {\em Chandra\/} results, 
despite the fact that these satellites were launched $5-7$ years earlier.
The angular resolution of these surveys limits the fluxes of the
optical counterparts to $R<21$~mag for the {\em BeppoSAX\/}
HELLAS survey and $R<19$~mag for the {\em ASCA\/} Large Area Sky 
Survey, in order to avoid confusion and speed
up the identification process. This is about two magnitudes brighter
than the {\em ROSAT\/} Deep Survey limits. At these levels, 
$\sim85$\% of the sources are identified with fairly high confidence.

The nature of the sources is rather different from the {\em ROSAT\/} 
sources, with approximately one-third not having prominent broad lines 
(La Franca et al.\ 2002; Akiyama et al.\ 2003),
while in the bright and faint {\em ROSAT\/} samples
(Lehmann et al.\ 2000; Appenzeller et al.\ 2000), more than 90\% of 
the AGN identifications have broad lines. The hard X-ray surveys find very
few narrow-line Seyfert~1 galaxies compared to the large fraction in the
{\em ROSAT\/} surveys. There are only a very few sources at these 
hard X-ray flux levels that have ``normal'' galaxies as optical 
counterparts.

The range of optical--to--X-ray flux ratios is very large, 
$\pm2$ dex, which is much larger than in the {\em ROSAT\/} bands, 
$\pm1$~dex. It is believed that there are two main effects in 
the sample differences: 1) obscuration, and 
2) redshift. It has been noted in the deep {\em ROSAT\/}, 
{\em Chandra\/}, and {\em XMM-Newton\/} samples 
(see \S\ref{richardsecchandra})
that there is a strong correlation (Hasinger et al.\ 1999) 
between $R-K$ color
and optical magnitude, with the fainter optical counterparts of the
X-ray sources being systematically redder. Part of this is clearly
due to the increasing effect of dust on the rest-frame UV colors of
AGN (the same amount of dust is much more significant in the UV than 
in the optical, and, of course, it is the UV which gets redshifted 
into the rest-frame optical band for $z>1$ objects), and part
seems to be due to the relative dominance of starlight in many 
of the faint sources and the steep SED of stellar objects. 
The mean X-ray spectral indices of the objects flatten as the sources 
get fainter by about $\Delta \alpha\sim0.3$ over a factor of $10^3$ 
in flux (Ueda et al.\ 1999).

The properties of these sources indicate that a significant fraction
of them have highly absorbed spectra, but the absorption is not
``simple''. The popular ``leaky absorber model'' (Turner et al.\ 1997), 
which fits many of the high signal-to-noise {\em ASCA\/} AGN spectra, 
produces a significant signal for highly absorbed objects in the soft 
($0.5-2$~keV) band. Recently, results from many {\em Chandra\/} and 
{\em XMM-Newton\/} serendipitous fields are becoming available, which 
will allow a much larger database for the 
$F_X<10^{-13}$~ergs~cm$^{-2}$~s$^{-1}$ sources, but the 
solid angles covered will be too small to produce many sources in 
the $F_X\sim10^{-12}-10^{-13}$~ergs~cm$^{-2}$~s$^{-1}$ range,
for which good X-ray spectra can be obtained.

It is entertaining to note that the recent deep {\em Chandra\/} hard
X-ray surveys have more objects per pointing than the entire 
{\em HEAO-1\/} or {\em ASCA/BeppoSAX\/} data sets.

\subsection{Deep {\em Chandra\/} and {\em XMM-Newton\/} Hard 
X-ray Surveys}
\label{richardsecchandra}

Because the {\em ROSAT\/} Deep Surveys resolved most of the soft X-ray
background, much of the emphasis with {\em Chandra\/} and {\em XMM-Newton\/}
has been on hard X-ray surveys\footnote{For the most recent update on the
{\em Chandra\/} and {\em XMM-Newton\/} surveys, see Barcons (2003).}.  
It also seems that at the faintest fluxes reached by {\em Chandra\/}
($\sim10^{-16}$~ergs~cm$^{-2}$~s$^{-1}$ in the
$2-10$~keV band, corresponding to 10 counts in the deepest 
$2\times10^6$~s exposures), the relative fraction of objects which are 
AGN declines rapidly (Hornscheimer et al.\ 2003), indicating that  
one has reached the ``end of AGN-ness'', similar to what is seen in 
the deep radio surveys.

The major advantage of the {\em Chandra\/} data is that, to very faint
optical magnitudes ($I\sim28$), there is almost always an ``unique'' 
optical identification (or lack thereof, Barger et al.\ 2003a;
Koekemoer et al.\ 2004). 
Thus, as opposed to almost all other surveys (with the exception of 
radio data), the identifications are certain, and one does not need 
to rely on optical spectroscopy to confirm the identification. 

The nature of the hard {\em Chandra\/} sources has been rather 
surprising (Mushotzky et al.\ 2000; Barger et al.\ 2001; 
Alexander et al.\ 2001). As summarized in Barger et al.\ (2003b), 
less than 30\% of the optical counterparts have
strong broad lines or a non-thermal continuum, and many of the
other 70\% are pure absorption-line objects. Most of the light from
these sources is due to stars in both the optical (Barger et al.\ 2002)
and IR (Crawford et al.\ 2001) bands; even with {\em HST\/} images, 
the nuclei are almost invisible (Grogin et al.\ 2003; Cowie \&
Barger, this volume).  
Most of these objects have stellar luminosities near $L^*$ or brighter 
in the $K$-band. Given the very high areal density of the 
{\em Chandra\/} sources (more than 3000~deg$^{-2}$ at 
$F_X\sim10^{-16}$~ergs~cm$^{-2}$~s$^{-1}$ in the $2-8$~keV band), 
this makes these ``optically-dull'' (a nomenclature first used by 
Elvis et al.\ 1981) objects the most numerous class of AGN. 
Similar results are seen in the {\em XMM-Newton\/} surveys
(Barcons et al.\ 2002). Comparison of the optical colors of the 
{\em XMM-Newton\/} sources shows that only about half are 
consistent with the region used by the SDSS to find AGN 
(Richards et al.\ 2002).

The exquisite {\em Chandra\/} positions have allowed the detection of  
the X-ray--selected AGN population that (a) dominates the AGN numbers,
(b) are not found by standard optical selection criteria, and (c) would 
not have been found in previous X-ray surveys. This is a stunning 
example of the ``lamppost effect'' alluded to in the introduction and 
should open our eyes to the possibility of even more such surprises 
when IR selection techniques are well developed.

\subsection{Comparison of X-ray and Optical Surveys}

So the question arises as to what is the difference between the X-ray
and optically-selected samples? Since, as discussed above, there are
strong selection effects in soft X-ray, UV, and optical samples, especially
at luminosities less than $10^{44}$~ergs~s$^{-1}$, it is not a great 
surprise to found that the major differences between these two surveys 
occur at this luminosity and less (Steffen et al.\ 2003). Clearly, 
optical-selection techniques are much less sensitive to low-luminosity
objects, where the effects of stellar dilution can be large 
(Moran et al.\ 2002; Moran, this volume).

Stellar dilution reduces the amplitude of optical
variability. In one of the very few direct comparisons of optical
variability versus X-ray selection, Dobrzycki et al.\ (2003) found a ratio
of X-ray to variability-selected quasars of $\sim5:1$ at $V<20.5$. Similar
effects are seen in a comparison of objects in the SSA13 field, where
Cowie et al.\ (1996) found only three obvious quasars, and a
moderately deep {\em Chandra\/} observation found $>20$ active galaxies 
in the same solid angle. Similar results are obtained in the Hubble Deep
Field-North (HDF-N) and Flanking Fields, where Liu et al.\ (1999) 
found an areal density of $B<21$ quasars of 30~deg$^{-2}$ (rather 
typical of optical quasar surveys), while there are $>1000$ 
{\em Chandra\/} AGN~deg$^{-2}$ in the same field.

Detailed analyses of several of the ``optically-dull'' galaxies
(including the famous first one, 3C264) show that dilution, while
important (Watanabe et al.\ 2002), is often not enough to remove an
obvious AGN signature (Severgnini et al.\ 2003), and the optical
continuum and emission lines must be reduced in strength if these
objects have the same ratio of H$\alpha$ or $B$-band flux to X-rays 
as ``normal'' AGN. A large fraction of the ``optically-dull'' galaxies
have hard X-ray spectra indicative of column densities
$>10^{22}$~atoms~cm$^{-2}$, but this is primarily derived from 
hardness ratios, rather than directly from X-ray spectral fitting.

Alternatively, it is possible that the ``optically-dull'' galaxies are
intrinsically weak in the optical band, as seen in many LINERs, which
have the same properties: high X-ray--to--optical ratios, absent or
very weak optical continua, no broad lines, and weak total 
emission-line flux (Ho 1999). This reviewer suspects that all three 
effects are important but that many of the {\em Chandra\/} 
``optically-dull'' galaxies
are higher luminosity versions of LINERs, which are not cataloged in the
low-redshift universe, since such objects would probably only be found in
the low-redshift universe via a large solid angle sensitive hard X-ray
survey. There are recent indications from the SDSS that such objects 
are quite common (as indicated in the Ho et al.\ 1995 work), but without 
X-ray measurements, their true luminosities are difficult to estimate.

Because of the absence of measurable optical nuclear light in these
objects, the bolometric correction factors are not known, and thus the
contribution of these objects to the mass density of black holes is
difficult to estimate (see, however, Cowie \& Barger, this volume). 
The recent {\em HST\/} observations of {\em Chandra\/} 
sources (Grogin et al.\ 2003) indicate that the observed nuclear light 
is $\sim20$ times less than anticipated on the basis of the X-ray flux 
and the X-ray--to--optical ratios of the {\em ROSAT\/} sources.

One of the unexpected features of the distribution of the {\em Chandra\/}
sources was their strong concentration in large scale structures
(Barger et al.\ 2003b; Gilli et al.\ 2003; Yang et al.\ 2003), in
contrast to optically-selected samples, which have the same correlation
functions as normal galaxies. I suspect that this can be understood
as a matter of the higher space density of X-ray--selected AGN that
allows them to be traced on smaller length scales not possible with 
optical samples.

Theoretically, this result is perhaps unexpected. It is believed that
there is a strong correlation between the mass of the black hole and
the mass of the galaxy. Therefore, luminous AGN should be in massive
galaxies, which are more strongly clustered than other galaxies.
However, the mean bolometric luminosity of the {\em Chandra\/} AGN sources 
is considerably less than that of objects in optically-selected surveys,
especially at $z>0.2$. Thus, one might naively expect X-ray--selected 
AGN to be more weakly correlated than optically-selected objects.

It is likely that the X-ray luminosity is strongly correlated with 
the black hole mass, with a scatter of roughly 50 (Grupe 2004).
This sort of correlation is apparently not 
seen with the bolometric luminosity (Woo \& Urry 2002). Thus, the 
correlation of X-ray luminosity with the optical luminosity of the 
host galaxy seen in the {\em Chandra\/} fields 
(Fig.~22 in Barger et al.\ 2003b) indicates that a very large
fraction of the ``medium luminosity'' {\em Chandra\/} AGN lie in 
massive galaxies that are strongly correlated
(Barger et al.\ 2001; Cowie et al.\ 2004b). I believe 
the true question should be: why does the optically-selected sample 
not show a stronger correlation function?

\subsection{Very High Energies}

The detection of luminous $E>10$~keV emission from several nearby,
apparently low-luminosity objects (e.g., NGC4945; Done et al.\ 2003)
shows that even the $2-8$~keV band can suffer from selection effects.
The {\em BeppoSAX\/} survey of Seyfert~2 galaxies 
(Risaliti, Maiolino, \& Salvati 1999) 
shows that the distribution of column densities is flat in log space 
from $10^{21}$ to $3\times10^{24}$~atoms~cm$^{-2}$.
While very high column density objects may be
absent from the {\em Chandra\/} samples, the argument put forward by Fabian
(2000) that three-fifths of the nearest AGN have very high column densities
has, in my opinion, strong merit. The absence of a large solid angle
sensitive $E>10$~keV survey means that the statistics of such objects
are hard to constrain, but it certainly seems possible that the number
of low to moderate-luminosity objects at $z<0.5$ with very large column
densities may be similar to that of less absorbed objects. The
fundamental question is whether ``absorbed'' accretion is a major
component to the total AGN luminosity. It is not clear how,
absent a new X-ray mission\footnote{It seems as if
{\em INTEGRAL\/} is not sensitive enough to perform a
survey of AGN in the $10-30$~keV band; the sensitivity of {\em SWIFT\/} 
is not yet clear, but it may provide the best hard X-ray survey yet 
performed.}, to proceed to answer this question.

\section[Conclusions]{Conclusions}

The wide variety of techniques to detect AGN has resulted in a vast
array of objects with an enormous variety of selection effects. At
high luminosities, optical and X-ray selection seem equally sensitive.
At lower luminosities, beam dilution, absorption, and the possible
existence of optically-quiet AGN seem to make X-ray selection more
reliable. While it may be premature, it seems as if hard X-ray
emission is a fundamental observational property of AGN and that the
most unbiased samples of AGN are found in $E>2$~keV X-ray surveys.
This field is rapidly changing, and I anticipate that we will soon
have direct comparisons of very sensitive techniques in the X-ray,
optical, radio, and IR bands in exactly 
the same places in the sky to allow the first full view of the 
AGN phenomena.

\begin{acknowledgments}
I would like to thank my collaborators on the {\em Chandra} Large Area 
Sensitive Survey, Len Cowie, Amy Barger, Aaron Steffen, and Yuxuan 
Yang for many exciting interactions over the last 3 years.
\end{acknowledgments}

\end{document}